\documentclass[]{copernicus}
\nolinenumbers

\begin{document}

\title{Super Resolution of Arctic Sea Ice Concentration}


\Author[1]{Jun}{Zhai}
\Author[1]{Cecilia M.}{Bitz}

\affil[1]{Department of Atmospheric Sciences, University of Washington, 98195, USA}




\runningtitle{Sea Ice Super Resolution}

\runningauthor{Zhai and Bitz}

\received{}
\pubdiscuss{} 
\revised{}
\accepted{}
\published{}


\firstpage{1}

\maketitle

\begin{abstract}
Arctic sea ice concentration is often coarsely observed and numerically computed despite its importance for polar climate system. In this work we present three machine-learning methods to recover the original high-resolution images from the coarse-grained low-resolution counterparts. The promising results indicate a possibility of extending the application to a broad range of geophysical variables.
\end{abstract}


\introduction 
As one of the critical parameters in the climate system, Arctic sea ice concentration has been widely monitored via polar orbiting satellites, using optical and passive microwave imagery and synthetic aperture radar. Gridded products at a regular spatial and temporal interval from these observations all share uncertainties owing to limited spatial and temporal coverage and/or coarse resolution. Measurements from multiple satellite overpasses are usually composited to produce gridded fields with complete spatial coverage. 
To improve the resolution obtained from a low sampling rate, measurements from multiple satellite overpasses are usually combined to generate a composited high-resolution estimate \citep{drue2004high,fraser2010generation,meierstewart2020}. The outcome of the compositing process, however, can produce a smoothed view of the original swath resolution owing to sub-scale averaging and interpolation that deviate from the truth at the scale of the composite.

Similar to satellite measurements, numerical models also face the challenge of resolving important features with low sea ice concentration, such as leads and polynyas, due to expensive computational costs. A tactic to mitigate computing time by taking longer time steps and employing unrealistically large diffusion required for numerical stability often spuriously smooths out features that are partially resolved. Processes that involve these unresolved features must be parameterized in dynamical models by predicting the influence of high-resolution on the resolved scales based on information at the resolved scale.

Super resolution (SR) is a strategy to mitigate information loss at small scales in satellite products and to construct faithful parameterization for numerical models. The SR technique is explored in this paper in order to reconstruct a high-resolution sea ice concentration given its low-resolution counterpart. Specifically, we test three types of deep learning models that have been shown to achieve a state-of-the-art success for SR in recent studies, which are convolutional neural networks (CNN), recursive convolutional network (RCN), and random forest-based regression model (RF-REG), and apply them to super-resolve coarse-grained sea ice concentrations.  

This paper is organized as follows. Section 2.1 reviews the relevant SR models in literature; section 2.2 explains the coarse-graining process we apply to a high-resolution image in order to obtain its low-resolution counterpart; section 2.3 presents three state-of-art SR models from the literature. These models produce high resolution reconstructions relative to a baseline reconstruction of simple bilinear interpolation. Section 3 presents the results of applying the three state-of-art models to sea ice concentration followed by Section 4 as a conclusion.

\section{Methodology}
This section first gives a broad briefing of SR methods (Section 2.1), then explains the procedure through which low-resolution data are obtained (Section 2.2), and lastly focuses on the three types of models developed to super resolve the Arctic sea ice concentration (section 2.3). 

\subsection{SR Background}
Over the past two decades, SR, the process of obtaining high resolution images from their low resolution counterparts, has been widely and well researched in a variety of fields, such as satellite imaging, medical image processing, facial image analysis, text image analysis, sign plates reading, biometric recognition, etc. Based on the context of applications, SR techniques can be grouped into a broad taxonomy, as suggested in Fig.1 in \cite{nasrollahi2014super}. 

The SR problem we aim to solve in this work is to reconstruct the high-resolution field of a geophysical variable given its low-resolution counterpart. Specifically, we have a single low-resolution input image and output its corresponding high-resolution output as a result of the reconstruction. Thus, according to \cite{nasrollahi2014super}, our problem falls into the category of a "Single Image" in a "Spatial Domain". The single-image SR problem is inherently "ill-posed" because the solution is not unique given a single low-resolution image. To mitigate this issue, prior knowledge is needed to constrain the solution space. In our case, the true high resolution fields for the training set must be available to train the SR model, which once trained can then be used to reconstruct the high-resolution field when it is unavailable.

To distill the prior, a state-of-the-art strategy is the example-based learning \citep{freeman2002example} which extracts the similarities among the sub-patches of a set of image and learns the mapping between the low and high resolution of these patches \citep{yang2014single}. Using sparse-coding method \citep{yang2010image} as a representative of example-based learning methods, \cite{dong2015image} showed that the pipeline of example-based learning is equivalent to a CNN, which directly learns an end-to-end mapping between low- and high-resolution images with little manual pre- or post-processing. In this spirit of end-to-end mapping, a variety of machine learning techniques have been proposed and shown to reach the state-of-the-art efficiency in different contexts, such as RCN \citep{kim2016deeply} and RF-REG \citep{schulter2015fast,dou2018medical}.

While the aforementioned published works mostly focus on public benchmark image data sets, SR in climate-related work is not new. For instance, \cite{keating2012new} and \cite{keating2015upper} used an empirical stochastic parameterization of ocean turbulence within a dynamical model to super resolve the ocean velocity and sea surface temperature given coarse-resolution observations. \cite{bolton2019applications} tested whether the unresolved subgrid turbulent processes can be revealed by low-resolution model data using CNN.

\subsection{Data processing}

In this paper, we aim to reconstruct finer-scaled sea ice concentrations from low-resolution fields. In practice, we utilize two years of daily sea ice concentration fields from a high-resolution dynamical sea ice model to train and evaluate the ML models. These are the high-resolution fields that our SR model is meant to reconstruct. To resemble sparse satellite measurements or low-resolution model output that we might want to super-resolve, we first coarse-grain the high-resolution dynamical model output by employing a two-dimensional spatial filter that is a simple average around a given grid cell $(i, j)$ with a downgrading factor $m$. Specifically,
\begin{equation}
c_L(i,j)=\frac{1}{(m+1)^2}\sum_{k=-m/2}^{m/2} \ \sum_{l=-m/2}^{m/2} c_H(i+l,j+k),
\label{eq:coarsen}
\end{equation}
where $c_L(i,j)$ is the low-resolution field and $c_H(i,j)$ is the high-resolution field. Each high-resolution grid cell appears exactly once in an average, so the size of the low-resolution field is reduced by $m+1$ in each direction. For example, downgrading by a factor of four ($m = 4$) averages 5x5 grid cells at high resolution, i.e. starting from 2 pixels away from the center to the left to 2 pixels away from the center to the right, into 1 grid cell at low resolution.
We present results with $m=4$ in Section 2, and explore various downscaling factors in Section 3.  


In addition to coarse-graining, we adopted the traditional patch-based operations \citep{ram2013image}, which has been shown to be more effective with SR at preserving the local texture of an image without being constrained on a single pixel. In the context of sea ice concentration, for example, patches as samples differ from each other due to different land/sea geography. Each high-resolution field is broken into $P$ non-overlapping patches prior to coarse graining. Hence, the low-resolution counterparts are also on patches, and the input to the ML model is $P$ times the number of daily fields. 

The sea ice concentration we use is from a historical simulation of the Community Earth System Model Version 2 (CESM2) with CICE5 as the sea ice component \citep{danabasoglu2020community}. The resolution of the sea ice in the Arctic is approximately 5 km. We divided the Arctic domain (north of 48 N) into 1067 patches of $32$x$32$ grid cells, each of which has at least 10\% coverage by non-zero sea ice concentrations based on an average of a year. 

We acquired daily fields from CESM2 in 2007 and 2008, for a total of 778,910 patches. Among these patches, the last 64,020 patches (about 10\%) are used for validation and testing, 32,010 for each. 

\subsection{Model architectures}
\begin{figure*}
\centering
\noindent\includegraphics[scale=0.45,trim={0cm 0cm 0cm 0cm}]{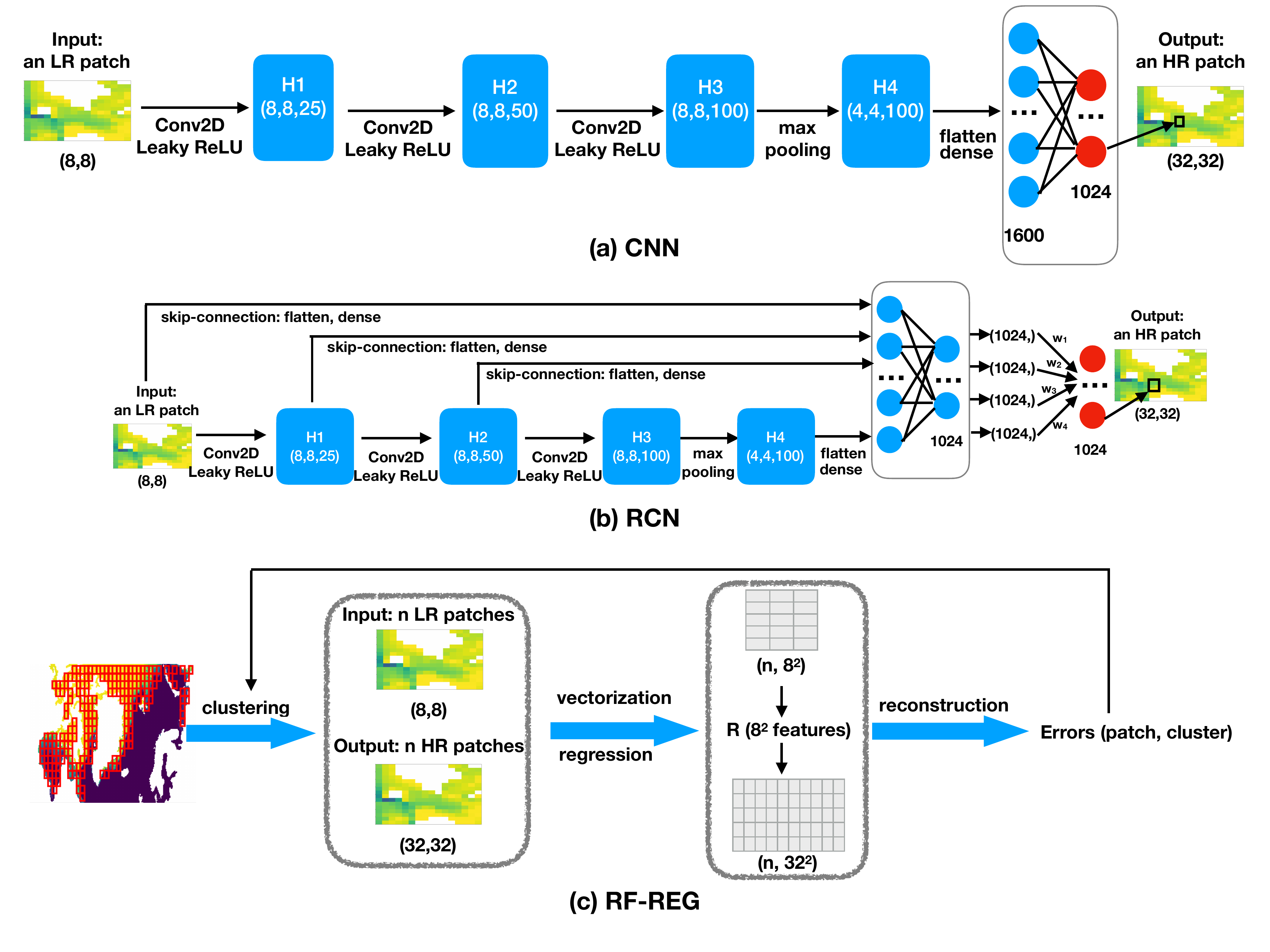}
\caption{The architectures for (a) CNN, (b) RCN, and (c) RF-REG with $m$=4.}
\label{f:arcs}
\end{figure*}

Distilled from the broad research literature aforementioned in Section 2.1, three representative ML methods are selected for this study to reconstruct the high-resolution Arctic sea ice concentrations. Specifically, We create SR models with CNN, RCN, and RF-REG, which are illustrated in Fig.\ref{f:arcs} and documented in detail below. The trainings for both CNN and RCN are implemented in Python with the Tensorflow/Keras library (https://keras.io). The training for RF-REG uses Python Scikit-Learn 0.24.2 library (https://scikit-learn.org). 

\begin{figure*}
\centering
\noindent\includegraphics[scale=0.45,trim={0cm 0cm 0cm 0cm}]{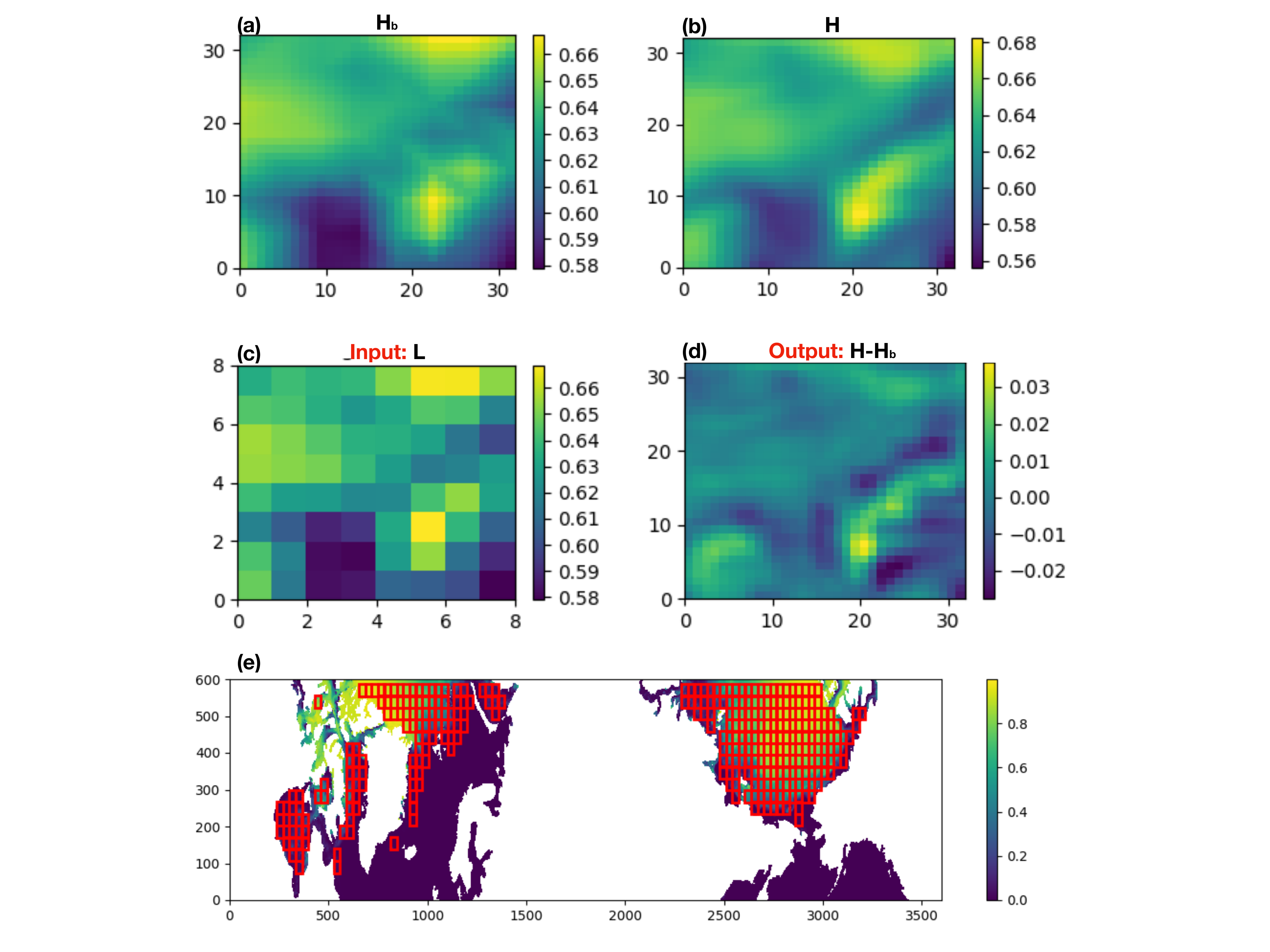}
\caption{Sample input and output with $m$=4 for our SR models. (a) $H_b$, the bilinear interpolated high resolution as baseline. (b) $H$, the original high resolution. (c) $L$, the coarse-grained low-resolution as input. (d) $H-H_b$, the difference between $H$ and $H_b$ as the prediction target. (e) Sea ice concentration patches of 32$\times$32 selected (red squares) over the Arctic.}
\label{f:IO}
\end{figure*}

The baseline estimate we define for each SR model to beat is the high-resolution reconstructed fields acquired from bilinear interpolation of the low-resolution fields. In fact,
we design the SR models to input low-resolution patches and output the difference between the high-resolution patch and the baseline bilinear interpolated patch (e.g., see Fig.\ref{f:IO}).

\subsubsection{CNN}
Convolutional neural network (CNN) is a class of deep-learning neural network that has been widely applied to image analysis. It breaks down an image composed of complex patterns into smaller and simpler pattern feature blocks by employing a mathematical operation called ``convolution". When $m$=4,
 the CNN model (Fig.\ref{f:arcs}(a)) takes a low-resolution image patch of size $8$-by-$8$ as input and then process it through three consecutive layers repeating a block unit: a two-dimensional (2D) convolutional layer and a LeakyReLU (Leaky Rectified Linear Unit) layer. The output of this three-time repeated block unit is then passed to a 2D max-pooling layer before getting flattened to a one-dimensional (1D) vector with a length of 1600. The final step is to regress the flattened 1D vector to another 1D vector of a size 51200, which is the flattened high-resolution 32$\times$32 patch.

\subsubsection{RCN}
To alleviate the issues of vanishing/exploding gradients with minimal redundant recursions, \cite{kim2016deeply} introduced a ``deeply-recursive convolutional network (RCN)" which recursively connects intermediate outputs to the terminal layer through ``skip-connections". Following this work, we construct an RCN adapted from the CNN architecture, as shown in Fig.\ref{f:arcs}(b). In the RCN architecture, intermediate predictions at each layer are output through a skip-connection and simultaneously supervised with respect to the truth to learn the optimal weights for each prediction. RCN resembles CNN for the basic blocks but adds the skip-connections which flatten the outputs from intermediate layers, dense them to 1D vectors of size 1024 as individual predictions, and then weigh these predictions via a supervised learning with respect to the truth. 

\subsubsection{RF-REG}
\cite{dou2018medical} proposed a random forest classifier that selects a regressor that matches a low-resolution patch to its counterpart in the high-resolution space. We name this method ``Random Forest-Regression" (RF-REG). 
The training algorithm is shown in Fig.\ref{f:arcs}(c) and summarized as follows. 
\begin{itemize}
\item Part I: Learn multiple regression models
    \item[] Step 1: Randomly and evenly divide all the training samples/patches into $j$ classes
    \item[] Step 2: Construct $j$ Ridge regression models with  Tikhonov regularization that map the low-resolution images to their high-resolution counterparts. 
    \item[] Step 3: Reconstruct all training samples and calculate the reconstruction errors using each of $j$ regression models respectively. 
    \item[] Step 4: Regroup the training samples according to the reconstruction errors calculated from Step 3.
    \item[] Step 5: Repeat Step 2 until the reconstruction errors converge.
\item Part II: Train an RF to classify which of the $j$ regression models in Part I a given patch belongs to. 
\end{itemize}
As a result, when a new low-resolution patch comes in, we pass it directly to the RF to decide which regression model to use for the high-resolution reconstruction. In this work we apply a Ridge regression with a regularization strength of 1.0 and an RF classification with the maximum depth of the tree being 2.

\section{Results}
\begin{figure*}
\centering
\noindent\includegraphics[scale=0.65,trim={4cm 4cm 4cm 6cm}]{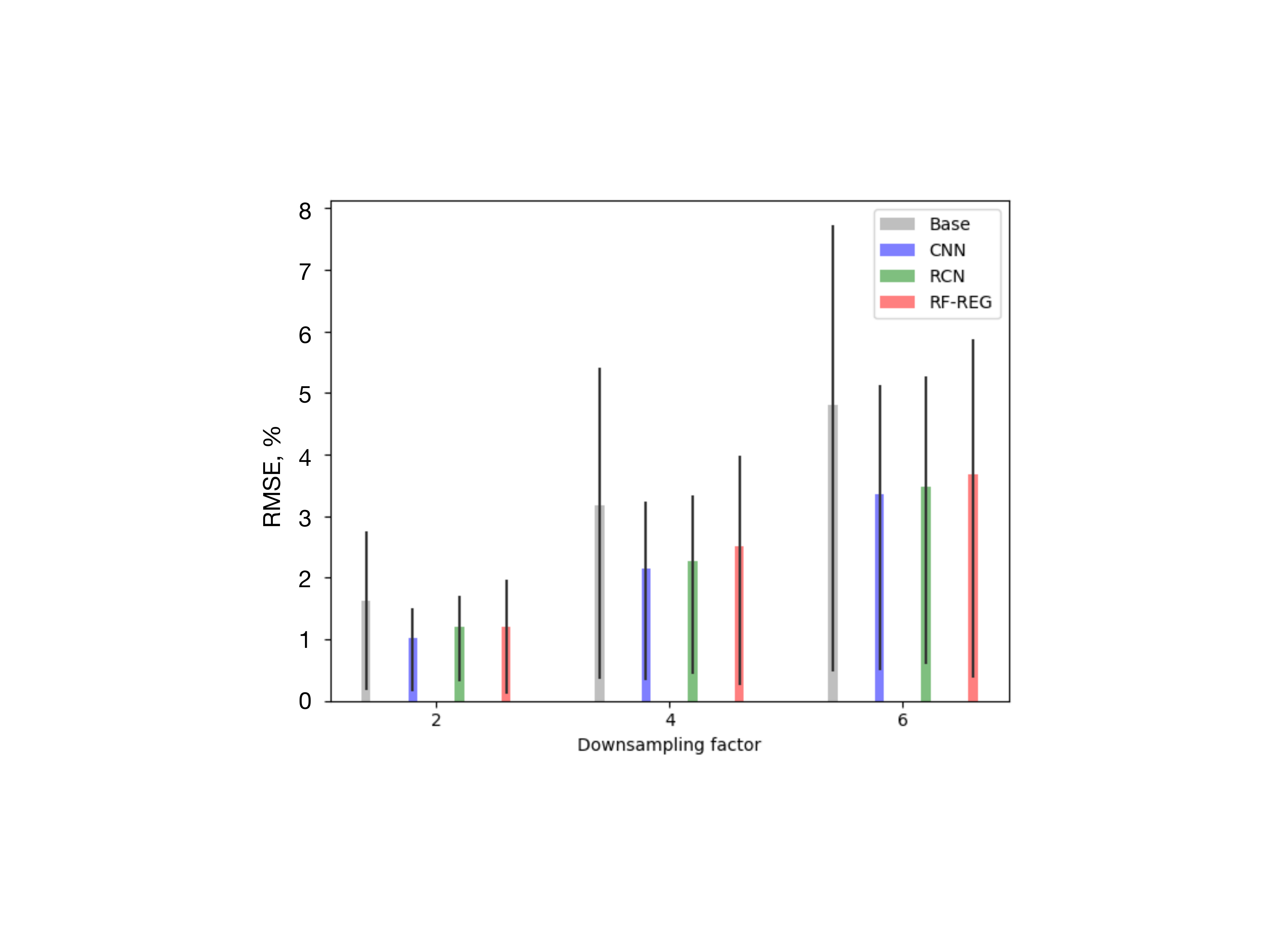}
\caption{The average RMSEs in \% of predictions by the baseline (gray), CNN (blue), RCN (green) and RF-REG (red) on the test set grouped by downsampling factors of 2, 4 and 6. Since the distribution of RMSE is not normal and skewed towards smaller values, we use top and bottom 25th quantiles to represent the spread of the average (black vertical lines). The computation of RMSEs does not include land or non-ice covered ocean.}
\label{f:compare}
\end{figure*}

We trained the three SR models based on ML methods each with three downsampling factors ($m=2,$ 4, and 6). The baseline estimates from bilinear interpolation plus the ML methods yield a total of 12 reconstructions. The performance of each reconstruction is evaluated by root mean square error (RMSE) (Fig.\ref{f:compare}) with the original high-resolution CESM2 output as the "truth". Finally, we average over the cells with non-zero ice concentration of all samples for all times within the test period to arrive at a single RMSE for each reconstruction for comparison (Fig.\ref{f:compare}). It follows that reconstructions from all three ML-based models surpass the baseline at each downsampling factor, with CNN having the smallest RMSE. The reason why RCN does not surpass CNN might be because we use leaky ReLU as the activation function in both CNN and RNN. Leaky ReLU is known to rectify the vanishing gradient problem, suggesting it is an important factor for this problem and therefore the skip-connection component in RCN does not appear to offer any additional advantage. 

The RMSE increases with downsampling factor for all reconstructions, in general, indicating, unsurprisingly, that the lower the input resolution is, the harder it is to reconstruct the high-resolution fields. The relative improvement of the reconstructions from the three ML-based SR models compared to the baseline decreases with downscaling factor, suggesting that the benefit of ML over bilinear interpolation decreases for harder problems. However, the relative RMSE for CNN compared to the baseline is 0.7 even when $m=6$, which indicates that CNN is arguably worth the trouble.


\section{Conclusions}
In this work we present three machine-learning methods, i.e. CNN, RCN and RF-REG, in SR models to reconstruct a high-resolution sea ice concentration from the corresponding coarse-grained low-resolution field. Compared to a baseline estimate from bilinear interpolation, all three ML-based SR models show superior performance, with CNN being the best. In addition, when applied to SST, CNN still surpasses the baseline estimate, which implies the potential promise of our results when applying this technique to other geophysical variables in general.

We envision a possible application of the SR methods in this study to the development of parameterizations in low-resolution Earth system models where high-resolution fields are reconstructed from the resolved model state and used to quantify a sub-grid scale process. For sea ice, this might include parameterizations of sea ice growth from unresolved small-scale ice-free or thin-ice areas, such as small leads and polynyas.

Another potential application is to satellite observations when transmission of satellite imagery is rate limited, especially for instruments that resolve fine spatial scales and/or have numerous spectral bands per pixel. Images at high-resolution could be transmitted intermittently with more numerous coarse-grained low-resolution images in the interim. The high-resolution images could then be used to train a ML-based SR model for reconstructing high-resolution images from the more common low-resolution images. However, this application might require significant adaption of the SR concept. For example, if the input information is purely satellite tracks, then 1D Convolution will be used instead of the current 2D convolution network. Also, since satellite tracks collect data at different time instants, the time scale complexity should also be taken into account.













\competinginterests{Both authors declare no competing interests for this work.} 



\bibliographystyle{copernicus}
\bibliography{reference.bib}

\end{document}